# Electron acceleration by magnetosheath jet-driven bow waves

Terry Z. Liu[1,2], Heli Hietala[3,4,5], Vassilis Angelopoulos[5], Rami Vainio[4], and Yuri Omelchenko[6]

[1]Cooperative Programs for the Advancement of Earth System Science, University Corporation for Atmospheric Research, Boulder, CO, USA. [2]Geophysical Institute, University of Alaska, Fairbanks, Fairbanks, AK, USA. [3]Department of Physics, Imperial College London, UK. [4]Department of Physics and Astronomy, University of Turku, Finland. [5]Department of Earth, Planetary, and Space Sciences, University of California, Los Angeles, USA. [6]Space Science Institute, USA

**Abstract**

Magnetosheath jets are localized fast flows with enhanced dynamic pressure. When they supermagnetosonically compress the ambient magnetosheath plasma, a bow wave or shock can form ahead of them. Such a bow wave was recently observed to accelerate ions and possibly electrons. The ion acceleration process was previously analyzed, but the electron acceleration process remains largely unexplored. Here we use multi-point observations by Time History of Events and Macroscale during Substorms from three events to determine whether and how magnetosheath jet-driven bow waves can accelerate electrons. We show that when suprathermal electrons in the ambient magnetosheath convect towards a bow wave, some electrons are shock-drift accelerated and reflected towards the ambient magnetosheath and others continue moving downstream of the bow wave resulting in bi-directional motion. Our study indicates that magnetosheath jet-driven bow waves can result in additional energization of suprathermal electrons in the magnetosheath. It implies that magnetosheath jets can increase the efficiency of electron acceleration at planetary bow shocks or other similar astrophysical environments.

# 1. Introduction

Downstream of Earth's bow shock, localized cold fast flow enhancements characterized by high dynamic pressure, referred to as magnetosheath jets, are observed frequently (several per hour, Plaschke et al., 2018 and references therein). Magnetosheath jets are typically ~1 $R_E$ in size (e.g., Plaschke et al., 2016) and occur nine times more often downstream of the quasi-parallel bow shock (the angle between upstream magnetic field and the bow shock normal $\theta_{Bn} < 45°$) than downstream of the quasi-perpendicular bow shock ($\theta_{Bn} > 45°$) (e.g., Vuorinen et al., 2019). The widely accepted explanation for this is that the quasi-parallel bow shock is very unstable with many ripples on its surface (e.g., Karimabadi et al, 2014; Hao et al., 2017; Gingell et al., 2017). When the solar wind crosses such a locally tilted surface, it is less thermalized and less decelerated than in the surrounding areas, resulting in a localized downstream flow that is colder and faster than the ambient magnetosheath (e.g., Hietala et al., 2009; Hietala et al., 2013). Occasionally, magnetosheath jets form also due to upstream drivers, such as solar wind discontinuities (Archer et al., 2012) and foreshock transients (Archer et al., 2014; Omidi et al., 2016).

When magnetosheath jets impact the magnetopause, they disturb both it and the magnetosphere-ionosphere system. For example, they can compress the magnetopause and trigger magnetic reconnection (Hietala et al., 2018). Such compression can also excite eigenmodes of the magnetopause surface (Archer et al., 2019). The perturbation on the magnetopause surface can then result in compressional low frequency waves within the magnetosphere, ionospheric flow enhancements, and auroral brightening (e.g., Hietala et al., 2012; Archer et al., 2013; Wang et al., 2018).

When magnetosheath jets are fast enough, they can drive a bow wave or even a secondary shock. As a supermagnetosonic magnetosheath jet approaches the magnetopause, a secondary shock propagating sunward in the plasma frame can form (Hietala et al., 2009; 2012). When the relative speed between the jet and the ambient magnetosheath flow is also supermagnetosonic, a bow wave or a secondary shock can form at the leading edge of the jet. Such a bow wave has been identified by both simulations (Karimabadi et al., 2014) and observations (Liu et al., 2019a), and has been shown to accelerate ions and possibly electrons (Liu et al., 2019a). The ion acceleration was explained with the help of a single particle model as due to ion reflection at the bow wave. A similar ion acceleration process, though at a different setting, was revealed by Vlasiator simulations at the bow wave of a fast-moving flux transfer event (Jarvinen et al., 2018). The electron acceleration process at bow waves or shocks ahead of jets, however, remains poorly understood as it has yet to be determined and analyzed comprehensively.

The above-mentioned observations of particle acceleration by jet-driven bow waves suggest that jets could play an important role in particle acceleration in shock environments. This is in light of the fact that shock acceleration, although one of the most important particle acceleration mechanisms in space, planetary and astrophysical plasmas, is still not fully understood. For instance, the theoretical acceleration efficiency of quasi-parallel shocks is not large enough to explain observations (e.g., Lee et al., 2012; Masters et al., 2013; Wilson et al., 2016). It is possible that jet-driven bow waves could provide additional energization to particles accelerated by the quasi-parallel shock and thus increase its acceleration efficiency when its jet-filled surrounding environment is properly accounted for. Therefore, it is necessary to understand how jet-driven bow waves accelerate particles and eventually incorporate this acceleration theory in quasi-parallel shock acceleration models. In this study, we apply case studies using multipoint Time History of Events and Macroscale Interactions during Substroms (THEMIS) observations to investigate how electrons interact with jet-driven bow waves. In the accompanying paper, Liu et al. (2019c submitted to JGR) present a statistical study to further confirm particle acceleration by jet-driven bow waves.

## 2. Data

We used data from the THEMIS mission (Angelopoulos, 2008) in 2008-2011, during which TH-A, TH-D, and TH-E (~10 $R_E$ apogee) were often in the magnetosheath (Sibeck and Angelopoulos, 2008). We analyzed plasma data from the electrostatic analyzer (ESA; 7 eV – 25 keV) (McFadden et al., 2008) and the solid state telescope (SST; 30 – 700 keV) (Angelopoulos, 2008) and magnetic field data from the fluxgate magnetometer (Auster et al., 2008).

Liu et al. (2019a) searched the event list reported by Plaschke et al. (2013) for magnetosheath jets and found 364 events (out of 2859) that have a bow wave or shock-like structure at their leading edge. The detailed selection criteria can be found in the accompanying paper, Liu et al. (2019c submitted to JGR). We selected three representative events that have electron energy flux enhancements associated with the bow wave for case studies.

## 3. Results

### 3.1. Overview

Figure 1 shows the overview plots of the three events on October 23, 13, and 24, 2011, respectively. Their solar wind conditions are listed in Table 1. In event 1 at ~14:02 UT, there was a fast magnetosheath jet (>300 km/s at ~14:02 UT in Figure 1.1c) with dynamic pressure larger than the solar wind dynamic pressure (Figure 1.1h). Ahead of the jet (yellow region in Figure 1.1), there were sharp increases in the magnetic field strength (Figure 1.1a) and density (Figure 1.1b), suggesting a bow wave. By using the coplanarity method and conservation of mass flux (Schwartz,

1998), we calculated the parameters of the bow wave (Table 1) showing that the fast-mode Mach number was ~1.4 ±0.2 (see calculation details in the supporting information).

This bow wave had likely steepened into a shock. When a cold fast flow supermagnetosonically compresses ambient hot plasma, there will be an interaction region hotter than both the fast flow and the ambient plasma, rather similar to a corotating interaction region. Just downstream of the bow wave, the interaction region can be seen with a wider ion distribution (white dashed circle in Figures 1.1d, e) than the ambient magnetosheath and the cold fast jet. In contrast, the interaction region was not observed for the bow wave reported in Liu et al. (2019a), possibly because its Mach number was only ~1.06 and thus the evolution was slower. Additionally, likely because the bow wave was still evolving, the velocity downstream of the bow wave was gradually varying resulting in non-zero divergence (Figure 1.1c). Thus, only the sharp enhancement of field strength and density were used to characterize the bow wave region (yellow).

Next, let us consider the electrons in event 1. No electrons were observed above 30 keV around the bow wave (Figure 1.1f, electron energy flux was below the SST noise level). Figure 1.1g shows electron energy spectra from 7 eV to 25 keV (by ESA). We see that there was energy flux enhancement at hundreds of eV to several keV by a factor of ~2.3 (after divided by the density increase ratio) just downstream of the bow wave and the maximum energy that has energy flux above the ESA noise level (black line) increased from ~3 keV to ~7 keV. This indicates that there was moderate electron acceleration/heating associated with the bow wave.

In event 2, there were also increases in density and field strength ahead of a fast jet (yellow in Figures 1.2a, b) suggesting a bow wave. However, because the magnetic field in the ambient magnetosheath was very turbulent, the field strength increase was not as sharp as the density increase. Thus, the uncertainty of the calculated shock parameters was much larger than in events

1 and 3 (Table 1). As for the electron energy spectra, there were tens of keV electrons in the ambient magnetosheath prior to the event (~14:19-14:20 UT in Figure 1.2f). Near the bow wave (~14:21 UT), their energy flux became enhanced by a factor of 1.3 on average and the maximum energy that has energy flux above the SST noise level increased from a typical value of 150 keV prior to the event to 200 keV just upstream of the bow wave event (white line in Figure 1.2f), suggesting moderate acceleration/heating at the bow wave. After the bow wave, the electron energy flux decreased. We will demonstrate why the electron energy flux increased near the bow wave and decreased after it in Sections 3.2 and 3.3.

In event 3, the bow wave with field strength and density enhancement can be seen at ~17:38:15 UT (yellow in Figures 1.3a, b). Right ahead of the bow wave, there were large amplitude magnetic fluctuations that are likely magnetosonic waves (Figure 1.3a). Similar to event 2, there were also tens of keV electrons in the ambient magnetosheath (~17:34 to 17:36 UT in Figure 1.3f). The energy flux enhancement near the bow wave (~17:36 to 17:38 UT) was more significant than in event 2, and the maximum energy increased from 150 keV to 300 keV (white line in Figure 1.3f). Next, we focus on this event exhibiting the most pronounced electron enhancement to investigate whether the enhanced electron energy flux was caused by the bow wave and what the acceleration process was.

### 3.2. Analysis of Event 3

This event was observed by three THEMIS spacecraft (see spacecraft position in Figure 2). TH-A and TH-E were very close to each other (~1000 km apart), and TH-D was ~4000 km and ~3000 km away from TH-A and TH-E, respectively (see Figure S1 in the supporting information for TH-A and TH-D observations). As a result, the calculated parameters of the bow wave by TH-A and TH-E were very similar to each other but different from those by TH-D (Table 1). Based on the bow wave normal directions we obtained at the three spacecraft, we estimate its scale size to be ~1 $R_E$, consistent with the typical size of magnetosheath jets previously reported in the literature (e.g., Plaschke et al., 2016). We sketch it accordingly in Figure 2.

Based on the geometry of the event (Figure 2), we propose the following hypothesis of the acceleration process: In the ambient magnetosheath, there were suprathermal electrons moving inside a flux tube (~17:34 to 17:36 UT in Figure 1.3f, orange region in Figure 2a). As the bow wave approached (black curve), it provided further acceleration, such as shock drift acceleration (red region in Figures 2b, c).

To support this hypothesis, we first need to confirm that the enhanced energy flux (~17:36 to 17:38 UT in Figure 1.3f) was indeed from the bow wave. To demonstrate the direction of electron motion, we compare the electron energy flux parallel and anti-parallel to the magnetic field. Because the bow wave was neither a tangential discontinuity (total pressure was not balanced and there was finite net flow across it) nor a perpendicular shock (Table 1), there was a continuous magnetic normal component across it. Because the magnetic field $B_x$ was overall positive (grey shading in Figure 3a) and the bow wave normal was mainly earthward (Table 1), the magnetic normal component was -1.2±0.3 nT pointing from upstream to downstream. As a result, anti-parallel (parallel) direction upstream (downstream) of the bow wave corresponds to a direction away from the bow wave.

Let us first consider electrons above 30 keV (i.e., within the SST energy range). Figure 3e shows the ratio of parallel flux to anti-parallel flux and Figure 3f shows its 9s-smoothed line plot by averaging over the first six SST energy channels from ~30 keV to 140 keV. We see that in the ambient magnetosheath (~17:34 to 17:36 UT), these suprathermal electrons were dominated by parallel (sunward) flux (blue in Figure 3e). When the spacecraft approached the bow wave, the anti-parallel (earthward) flux started to dominate (red in Figure 3e). This may indicate that the enhanced electron energy flux came from the bow wave. After the spacecraft crossed the bow wave (vertical dashed line in Figure 3), the parallel (sunward) flux dominated (blue in Figure 3e). This trend can be more clearly seen in Figure 3f: the smoothed ratio of parallel flux to anti-parallel flux crossed the value of one at the vertical dashed line. Such bi-directional flux away from the bow wave further suggests that the enhanced electron energy flux in Figure 3b could be from the bow wave (two red arrows in Figure 2b). Later, we will further examine the reason of such anti-parallel/parallel anisotropy.

With regard to electrons below 30 keV (measured by ESA), Figure 3j shows the ratio of their parallel to anti-parallel flux. In the ambient magnetosheath (~17:34 to 17:36 UT), we see that there were multiple populations (separated by horizontal dashed lines in Figure 3j): Electrons below 20 eV were dominated by anti-parallel anisotropy (red). These were earthward moving magnetosheath thermal electrons. Electrons between 20 eV to 200 eV were dominated by parallel anisotropy (blue). Electrons between 200 eV to 2 keV were mostly anti-parallel (red). Above 2 keV, because the energy flux was close to the ESA noise level, only one (the lowest in that energy band) energy channel can be used. We see that electrons in that energy were mainly in the parallel direction same as those measured by SST. Later, we will demonstrate by smoothing the electron energy flux over time to lower the instrumental noise level that above 2 keV electrons behave consistently as one population.

When the spacecraft approached the bow wave (~17:36-17:38 UT), all the electron populations measured by ESA became mainly anti-parallel, i.e., they were moving earthward and away from the bow wave (red in Figure 3j). Downstream of the bow wave, electrons above 200 eV turned to be in the parallel (sunward) direction around 17:38:30 to 17:39:00 UT (blue) and electrons around

1 keV continued to be so until ~17:40 UT. This result is consistent with SST measurements at higher energies, showing that suprathermal electrons (>200 eV) were moving away from the bow wave on both sides. We thus confirm that the bow wave could be the energy source of electron energy flux enhancement.

Next, we discuss how the bow wave enhanced the electron energy flux by investigating electron phase space density (PSD) spectra (Figure 4). Figure 4a shows the averaged omni-directional phase space densities over time in the ambient magnetosheath (~17:34-17:36 UT; magenta line) and upstream of the bow wave (~17:36-17:38 UT; blue line). We see that there are multiple populations, corresponding to horizontal dashed lines in Figure 3h-j. Below 200 eV, electrons were probably a thermal population with a Maxwellian-like distribution. Between 200 eV and 2 keV, there was a suprathermal population following a power law distribution with a slope of ~5.1±0.2 (suprathermal 1). Above 2 keV, there was another suprathermal population also following a power law distribution but with a different slope of ~3.6±0.06 (suprathermal 2).

Next, we compare PSD spectra in the direction anti-parallel, parallel, and perpendicular to the magnetic field, respectively, to examine how they evolved from background magnetosheath to upstream and downstream of the bow wave (Figures 4b-g). The dashed lines are the omni-directional spectra as a reference to compare with spectra in three directions. We first investigate suprathermal population 2 measured by SST (Figures 4b-d). For electrons right upstream of the bow wave (between vertical blue line and dashed line in Figures 3c-e; blue in Figure 4), their PSDs in the anti-parallel, parallel, and perpendicular directions are larger than, weaker than, and similar to the omni-directional PSD, respectively (consistent with Figures 3c-f). Electrons in the background magnetosheath (between two vertical magenta lines in Figures 3c-e; magenta in Figure 4), on the other hand, have weakest PSD in the anti-parallel direction (corresponding to blue in

Figures 3c, e, f). As a result, the PSD enhancement from ambient magnetosheath to the upstream of the bow wave was dominant in the anti-parallel direction with ratio of ~4.6 (corresponding to energy increase ratio of ~1.5). This suggests that the acceleration was mainly in the anti-parallel direction. The moderate PSD enhancements in the other two directions were likely caused by the pitch-angle scattering from the anti-parallel direction possibly due to the magnetosheath turbulence or waves during and after the acceleration.

For electrons right downstream of the bow wave (between vertical dashed line and green line in Figures 3c-e), their parallel PSD was similar to that right upstream of the bow wave (green and blue in Figure 4c). Anti-parallel PSD (Figure 4b), on the other hand, decreased by ~50% from right upstream to right downstream of the bow wave (resulting in blue in Figures 3e, f). It is likely that as there was no particle source downstream of the bow wave and the anti-parallel electrons returned upstream, the anti-parallel PSD downstream of the bow wave can only decrease. This indicates that the particle source was from the upstream side of the bow wave.

Next, we examine electrons measured by ESA (Figures 4e-g). For thermal populations below 200 eV, there was no clear difference from background magnetosheath (between two magenta lines in Figures 3h-j; magenta in Figure 4) to upstream of the bow wave (between two blue lines in Figures 3h-j; blue in Figure 4). Further downstream of the bow wave (between two green lines in Figures 3h-j; green in Figure 4), the enhancement in PSD was due to the density enhancement. For suprathermal population 1 (200 eV to 2 keV) upstream of the bow wave, we see that anti-parallel PSD was larger than the parallel PSD (blue in Figures 4e, f by comparing with the dashed line, the omni-directional spectra; consistent with red in Figure 3j). In the background magnetosheath (magenta), such anti-parallel anisotropy was stronger (consistent with darker red in Figures 3j). As a result, the anti-parallel PSD enhancement was weaker than parallel PSD enhancement. One possible reason is that anti-parallel electrons was scattered to other directions (likely due to turbulence), as spectra upstream of the bow wave were more isotropic than the background. In an extreme case when electrons were perfectly isotropic upstream of the bow wave, the anti-parallel PSD enhancement would be always weaker than that in other directions, although the acceleration could be in the anti-parallel direction.

For electrons further downstream of the bow wave (between two green lines in Figures 3h-j), their PSD above 100s of eV in the parallel and perpendicular directions do not show clear difference compared to the background PSD (green and magenta in Figures 4f, g; similar colors in two regions in Figure 3i). In the anti-parallel direction (Figure 4e), on the other hand, PSD downstream of the bow wave shows clear depletion (resulting in blue in Figures 3h, j). This is consistent with SST results (Figure 4b) confirming that the particle source was from the upstream side of the bow wave.

Finally, we propose a possible acceleration mechanism based on our spectra plots, shock drift acceleration or the fast Fermi acceleration mechanism (e.g., Wu, 1984). The bow wave had a strong magnetic gradient. In the normal incidence frame, upstream electrons outside the loss cone can grad-B drift in the direction opposite to the convection electric field to gain energy and be reflected upstream. Such reflection with energy increase can result in the anti-parallel flux enhancement upstream of the bow wave. The energy increase is $2(mV^2/\cos^2\theta_{Bn} + mVv_\parallel/\cos\theta_{Bn})$, where $V$ is the magnetosheath flow speed in the normal incidence frame and $v_\parallel$ is the initial parallel speed of a particular electron (Krauss-Varban and Wu, 1989). As the local bow wave was nearly perpendicular ($\theta_{Bn} = 83 \pm 1.8°$), the energy increase was significant (e.g., $V = \sim 500\ km/s$, and if $v_\parallel = 10^4\ km/s$, the energy increase is ~700 eV). Electrons within the loss cone, on the other hand, crossed the bow wave. They could be shock-heated through the cross-shock potential but only for a few to tens of eV for low Mach number (e.g., Treumann, 2009; Cohen et al., 2019).

Meanwhile, anti-parallel electrons returned upstream resulting in the depletion in the anti-parallel flux. This acceleration process explains the "bi-directional" flux across the bow wave (Figures 3e, f, j; red arrows in Figure 2b)

This possible acceleration process, however, cannot maintain the spectra shape as shown in Figure 4. One possibility is that because turbulence can result in the power law spectra of electrons (e.g., Ma and Summers, 1998; Lu et al., 2011), it is likely that the magnetosheath turbulence continuously reshaped the electron spectra during and after the acceleration process just like in the background magnetosheath. This can explain why electron spectra have the same slope in different regions.

Figure 5 shows the comparison of the energy flux spectra between TH-E and TH-D (separated by ~ 3000 km; also see Figure S1 for detailed TH-D observations). We see that even though TH-D observed stronger background electron energy flux at ~17:35 UT than at TH-E (blue in Figures 5g, h), the energy flux enhancement near the bow wave at TH-E was stronger than at TH-D (~17:36-17:38 UT, red in Figures 5g, h). As a result, the anti-parallel PSD enhancement ratio from background magnetosheath to upstream of the bow wave observed by TH-D was ~2.7, smaller than ~4.6 observed by TH-E. This is consistent with that TH-E observed a larger $\theta_{Bn}$ than TH-D (Table 1), corresponding to stronger acceleration.

Downstream of the bow wave, the electron energy flux above 2 keV disappeared very rapidly (Figures 3b, g). We suspect that this is because the bow wave was curved, and the field lines were highly tilted downstream of the bow wave (see the zoomed in sketch in Figure 5i). The ambient electrons above 2 keV were in a flux tube of limited spatial scale. (As shown in the longer time interval in Figure S2 in the supporting information, such population was observed only for a short time.) Based on the observed time scale of this electron population (several minutes), its spatial scale was ~2-4 $R_E$ in GSE-Y (Figure 1.3c, $V_y$~100 km/s). The field lines downstream of the bow wave propagated at ~-200 km/s in GSE-X (Figure 1.3c). As the downstream field lines were very tilted, the spacecraft may need just ~10-20 s (several minutes · $V_y/V_x$ · $B_x/B_y$, where $B_x/B_y$ ~0.1) to pass through the entire particle source region connected to the bow wave.

Next, we discuss where suprathermal electrons in the ambient magnetosheath came from. Based on Figure 3e, we see that electrons above 2 keV were mainly in the parallel direction (sunward). One possible explanation is that because $B_z$ was negative, there could be magnetic reconnection at the magnetopause which caused suprathermal electrons to leak from the magnetosphere (the spacecraft was very close to the magnetopause seen in Figures 2 and S2). As for electrons below 2 keV, they were likely solar wind electrons heated/accelerated by the bow shock.

### 3.3 Analysis of Event 1 and 2

We apply similar analysis on events 1 and 2 (Figure 6). In event 1, the magnetic field was overall sunward (gray region in Figure 6a). Figure 6f shows the ratio of parallel flux to anti-parallel flux. Before and after the bow wave (vertical dashed dotted line), the suprathermal electrons above 200 eV were mainly moving in the anti-parallel (red; earthward) and parallel (blue; sunward) directions, respectively. . But different from event 3, in addition to the anti-parallel flux decrease downstream of the bow wave, there was also increase in the parallel flux corresponding to energy flux enhancement at ~14:02 UT (Figure 6e). The perpendicular flux normalized to omni-directional flux, on the other hand, only slightly varied (Figure 6g). The acceleration in the parallel/sunward direction downstream of the bow wave could be due to cross-shock potential (Krauss-Varban and Wu, 1989). When the spacecraft moved farther away from the bow wave after 14:02:40 UT, the electrons became earthward (red) again.

In event 2, the magnetic field was mainly earthward near the bow wave (gray region in Figure 6h). Because the energy flux measured by SST was not strong enough, the ratio of parallel to anti-parallel flux was very noisy. We only show the flux ratio measured by the ESA (Figure 6m). We see that the suprathermal electrons between 100 eV to 1 keV were mainly moving in the parallel direction (blue; earthward) before the bow wave (the first vertical dotted line) due to parallel flux enhancement (reflection). After the bow wave, in the downstream region (between two vertical dotted lines), the electrons were mainly moving in the anti-parallel direction (red; sunward) due to depletion in parallel flux (return upstream). After the spacecraft left the jet, the electrons turned back to being earthward. Therefore, the whole process in event 2 is consistent with event 3. The perpendicular flux also does not show any clear changes (Figure 6n).Similar to event 3, we also see that the electron energy flux measured by SST (Figure 6k) decreased rapidly across the bow wave. It may similarly be due to the very tilted magnetic field lines downstream of the bow wave and the spacecraft was quickly passing through the particle source region.

## 4. Conclusions and Discussion

In this study, we showed that magnetosheath jet-driven bow waves can further enhance the electron energy in the ambient magnetosheath. We summarize the observed process as follows: The spacecraft first observed suprathermal electrons in the ambient magnetosheath (Figure 2a). When the bow wave approached and the magnetic field lines connected to it, the spacecraft observed earthward enhanced electron energy flux from the bow wave (Figure 2b). After the spacecraft crossed the bow wave, depleted earthward electron flux was observed resulting in "bi-directional" motion across the bow wave (two red arrows in Figure 2b). The acceleration process is likely that when suprathermal electrons in the ambient magnetosheath cross the bow wave, some of them are energized through shock drift/fast Fermi acceleration while being scattered by magnetosheath turbulence. The rest of them continue moving downstream. Our results suggest that magnetosheath jet-driven bow waves could contribute to particle acceleration in the shock environment.

Other than the shock drift acceleration, there could be other electron acceleration mechanisms acting simultaneously, but their role was likely limited. The shock surfing mechanism is a possible acceleration process (Hoshino, 2001). However, as the bow wave Mach number is very weak, the theoretical energy increase is estimated as only 10s – 100s of eV (Treumanm, 2009 and references therein). This mechanism cannot explain the energy increase at 10s of keV in event 3, but could contribute in event 1 and 2. Additionally, as there was a local minimum magnetic field strength at ~17:37 UT in event 3 (Figure 3a) surrounded by the approaching bow wave and the other magnetic mirror at ~17:36 UT, electrons might experience Fermi acceleration by bouncing between them. The sunward anisotropy at ~17:36 UT in Figures 3e, f may indicate the reflection at the magnetic mirror. However, TH-D did not observe such a magnetic field configuration but a very small magnetic hole at ~17:37 UT (Figure S1). Therefore, the Fermi acceleration might contribute locally but was not the dominated process throughout the bow wave.

In the accompanying paper, Liu et al. (2019c submitted to JGR) employ a statistical study showing that magnetosheath jets that have a bow wave have a higher probability to exhibit higher electron energy than those without a bow wave. This shows that it is common for magnetosheath jets to accelerate electrons. The statistical study also shows that magnetosheath jets that have a bow wave can enhance the electron energy flux of ambient magnetosheath by a factor of 2 on average above ~100 eV. Such a result is consistent with our case study here, in Figure 4. Both the multi-case study and the statistical study show that electrons below ~100-200 eV do not have clear energy flux enhancement. One possible reason is that the cross-shock potential of the bow wave could complicate the motion of thermal electrons and prevent them from reflecting upstream.

Shock acceleration is one of the most important acceleration mechanisms in the universe. One of the most accepted shock acceleration mechanisms is the diffusive shock acceleration (e.g., Lee et al., 2012), i.e., particles bounce back and forth across the converging shock. While the bouncing particles are in the downstream region, they can be further energized by jet-driven bow waves. Here we estimate the contribution of bow waves to this process in the environment of Earth's bow shock. Based on fast Fermi model (Krauss-Varban and Wu, 1989), if the loss cone angle at the bow wave is 45° (e.g., event 3), 50% of incoming suprathermal electrons can reflect and gain velocity $2V/\cos\theta_{Bn}$. Because bow waves are mainly in earthward direction and magnetic field in the ambient magnetosheath dominates in YZ direction, $\theta_{Bn}$ is typically larger than 45° (Table 1). Additionally, fast wave speed in the magnetosheath is several hundred km/s and V should be faster than that to form a bow wave. Therefore, $2V/\cos\theta_{Bn}$ is typically around several thousand km/s (e.g., 8300 km/s in event 3). Based on the statistical study in Liu et al. (2019c submitted to JGR), the occurrence of jet-driven bow waves is ~0.2 per hour on average (depending on solar wind conditions). If we assume that each bow wave can exist and accelerate electrons for 1-2 min, the average velocity increase gained by electrons is 50%×0.2×1.5/60×thousands of km/s ~ a few to tens of km/s (e.g., 20 km/s in event 3). For diffusive shock acceleration, electrons gain velocity comparable to the velocity difference between the solar wind and magnetosheath for each bounce (e.g., Drury, 1983), which is typically several hundred km/s. Because each time electrons enter the magnetosheath, jet-driven bow waves could provide additional a few to tens of km/s on average, jet-driven bow waves could result in first-order modification (a few to ten percent) to the diffusive shock acceleration model. Under favorable solar wind conditions, such as high solar wind Alfvén Mach number (Liu et al., 2019c submitted to JGR), bow waves could contribute even more.

Upstream of shocks in the foreshock, foreshock transients can also drive secondary shocks and accelerate particles (e.g., Liu et al., 2016; Liu et al., 2017). Such secondary shocks can also further accelerate ambient suprathermal electrons in the foreshock to 100s of keV, similar to jet-driven bow waves observed in this study (Liu et al., 2019b). Nonlinear structures with secondary shocks/bow waves exist both upstream and downstream of the parent shock and both can accelerate particles contributing to the parent shock acceleration. Therefore, the shock environment is not just the shock itself but includes the multiple nonlinear structures surrounding it; those structures and their collective interaction should be included in future shock models.

**Table 1.** The solar wind dynamic pressure, solar wind cone angle, solar wind Alfven Mach number, solar wind plasma beta (corresponding to the solar wind conditions discussed in the accompanying paper, Liu et al. (2019c submitted in JGR)), the jet-driven bow wave normal, bow wave normal speed in the spacecraft frame, $\theta_{Bn}$, fast-mode Mach number of the bow wave, ambient magnetosheath plasma beta for three events. The uncertainty is obtained by varying the time interval used for parameter calculation (blue regions in Figure 1; see calculation details in the supporting information).

| Event # | SW Pdyn [nPa] | SW cone [°] | SW MA | SW Beta | normal | normal error [°] | Vshn in sc frame [km/s] | θ$_{Bn}$ [°] | Fast Mach | Beta |
|---|---|---|---|---|---|---|---|---|---|---|
| 1 | 3.2 | 33 | 10.6 | 1.9 | [-0.86, -0.48, 0.08] | 5.8 | 521±69 | 65±5.4 | 1.4±0.2 | 4.4±0.7 |
| 2 | 3.0 | 58 | 56 | 15 | [-0.82, -0.09, 0.39] | 21.7 | 181±50 | 86±5.1 | 0.93±0.33 | 6.2±0.6 |
| 3 THE | 0.8 | 22 | 6.9 | 1.0 | [-0.96, -0.22, 0.11] | 1.6 | 513±12 | 83±1.8 | 1.8±0.05 | 13±1 |
| 3 THA | | | | | [-0.98, -0.10, 0.07] | 3.5 | 436±35 | 72±5.5 | 1.5±0.1 | 13±3 |
| 3 THD | | | | | [-0.66, -0.35, 0.65] | 2.6 | 414±28 | 49±3.9 | 1.3±0.1 | 3.4±0.5 |

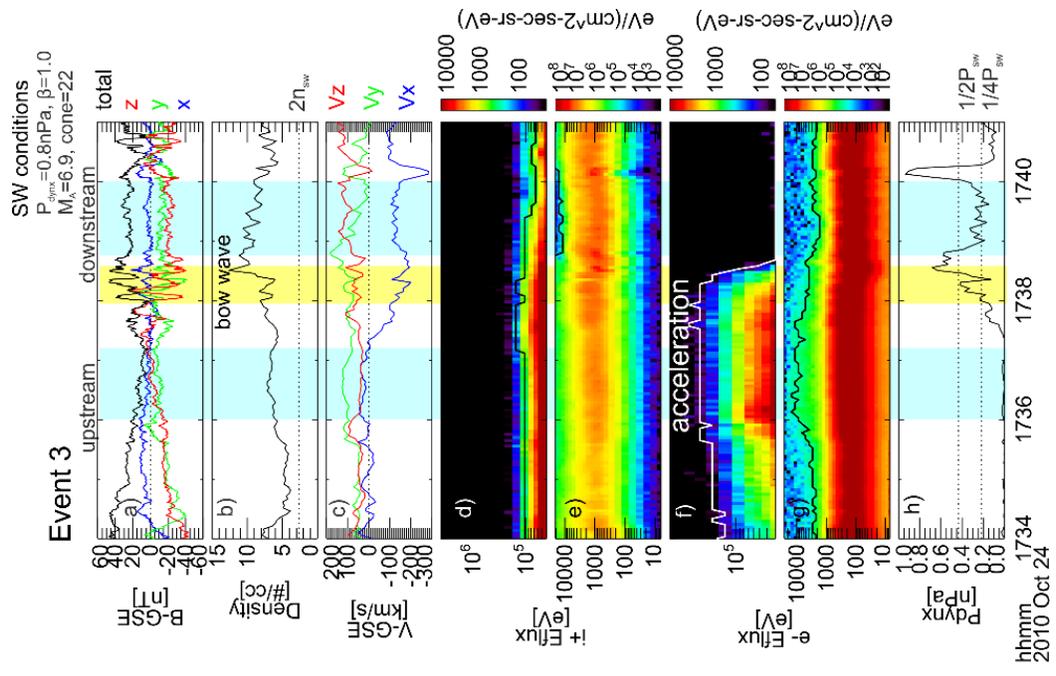
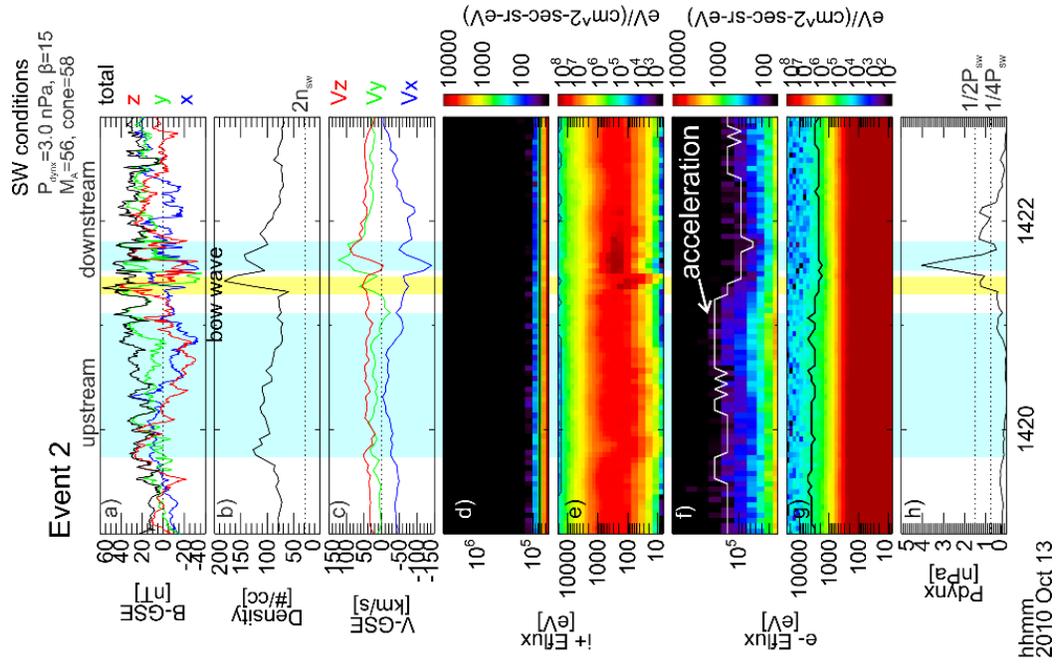
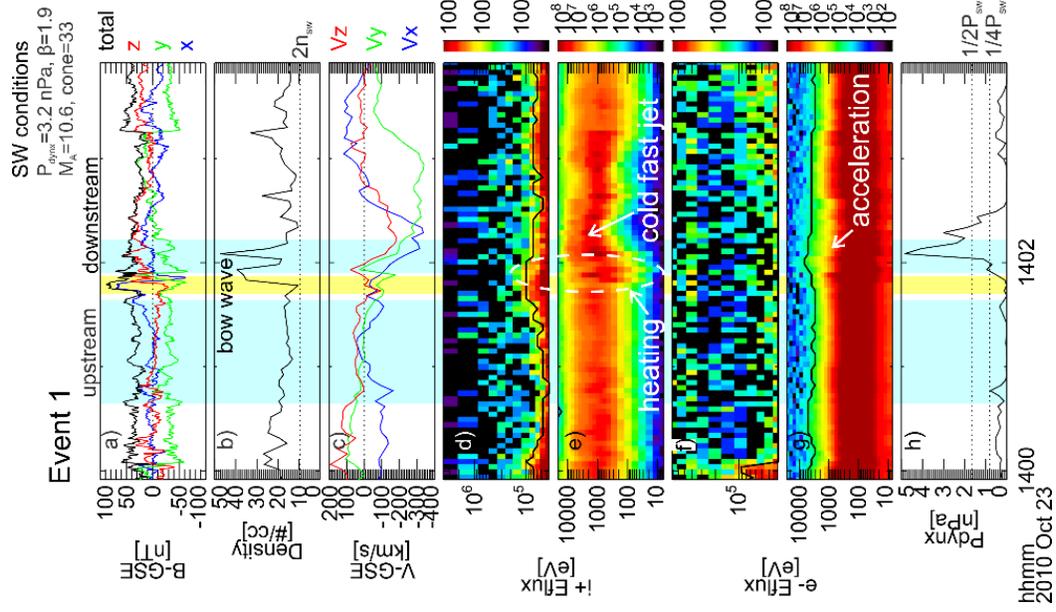

**Figure 1.** Overview of three events. Figure 1.1 (event 1) from top to bottom are TH-D observations of: (a) magnetic field in GSE; (b) density (the dotted line indicates two times the solar wind density); (c) ion bulk velocity in GSE; (d) ion energy flux spectrum from 30 keV to 700 keV; (e) ion energy flux spectrum from 7 eV to 25 keV; (f) electron energy flux spectrum from 30 keV to 700 keV; (g) electron energy flux spectrum from 7 eV to 25 keV; (h) dynamic pressure calculated using velocity in GSE-X component (two dotted lines indicate 1/2 and 1/4 solar wind dynamic pressure, respectively). The black lines in (d)-(g) represent the highest energy channel that has energy flux larger than the instrumental noise level. Figure 1.2 and 1.3 (event 2 by TH-A and event 3 by TH-E) are in the same format as Figure 1.1. Blue regions are the time interval used to calculate bow wave parameters.

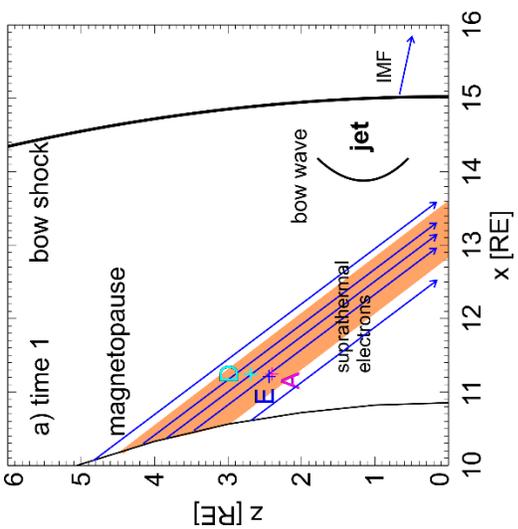
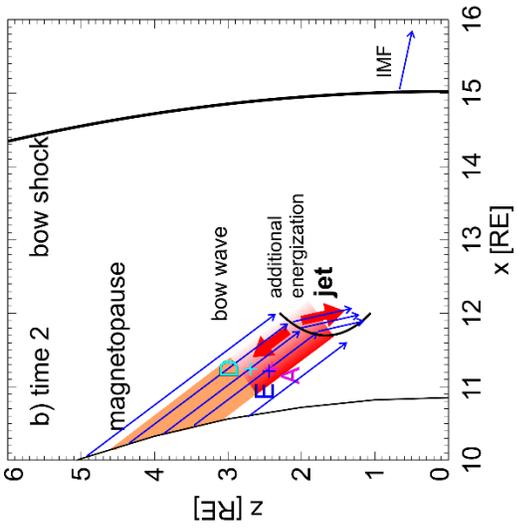
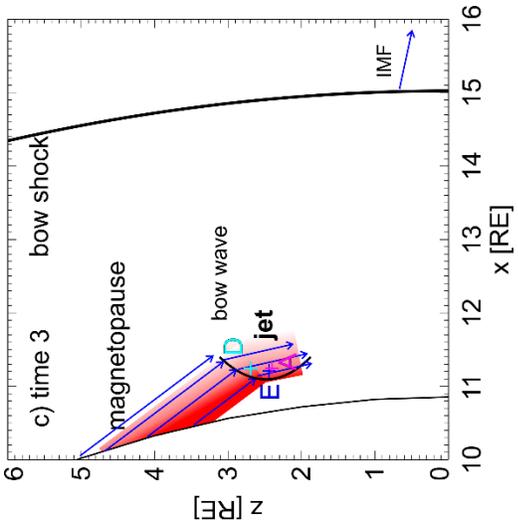

**Figure 2.** The sketch of event 3 indicating TH-A, D, E position (magenta, light blue, and dark blue crosses, respectively), relative to the magnetopause (from Shue et al. (1998) model) and the bow shock (from Merka et al. (2005) model). (a)-(c) show the earthward propagation of the bow wave (black curve) at three moments (time 1, 2, and 3). After the bow wave encountered the suprathermal electrons in the ambient magnetosheath (orange region), electrons were accelerated (red region) and streamed away from the bow wave (red arrows). The blue arrows indicate the magnetic field direction.

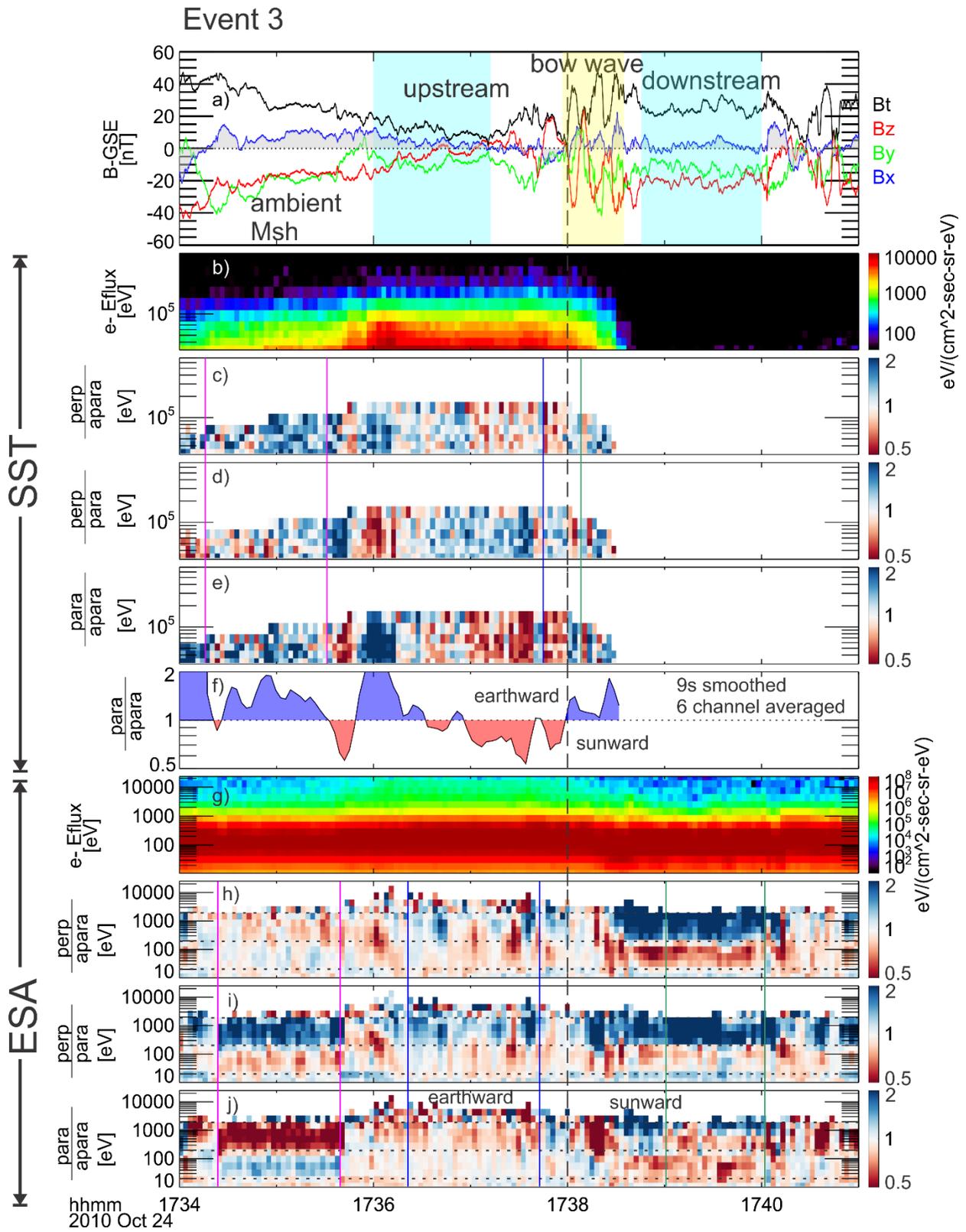

**Figure 3.** TH-E observations of electron anisotropy. (a) is the magnetic field in GSE and the shaded region indicates the sign of $B_x$ (the blue and yellow regions are the same as in Figure 1.3). (b) is electron energy flux spectrum from 30 keV to 700 keV. (c)-(e) are the ratio of perpendicular flux to parallel flux, perpendicular flux to anti-parallel flux, parallel flux to anti-parallel flux, respectively. (f) is the averaged value of (e) over the first 6 energy channels and 9s. (g)-(j) are the same format as (b)-(e) but from 7 eV to 25 keV. The vertical dashed line indicates the encounter of the bow wave. Colored vertical lines in (c)-(e) and (h)-(j) indicate the time interval of PSD spectra in Figures 4b-d and 4e-g, respectively.

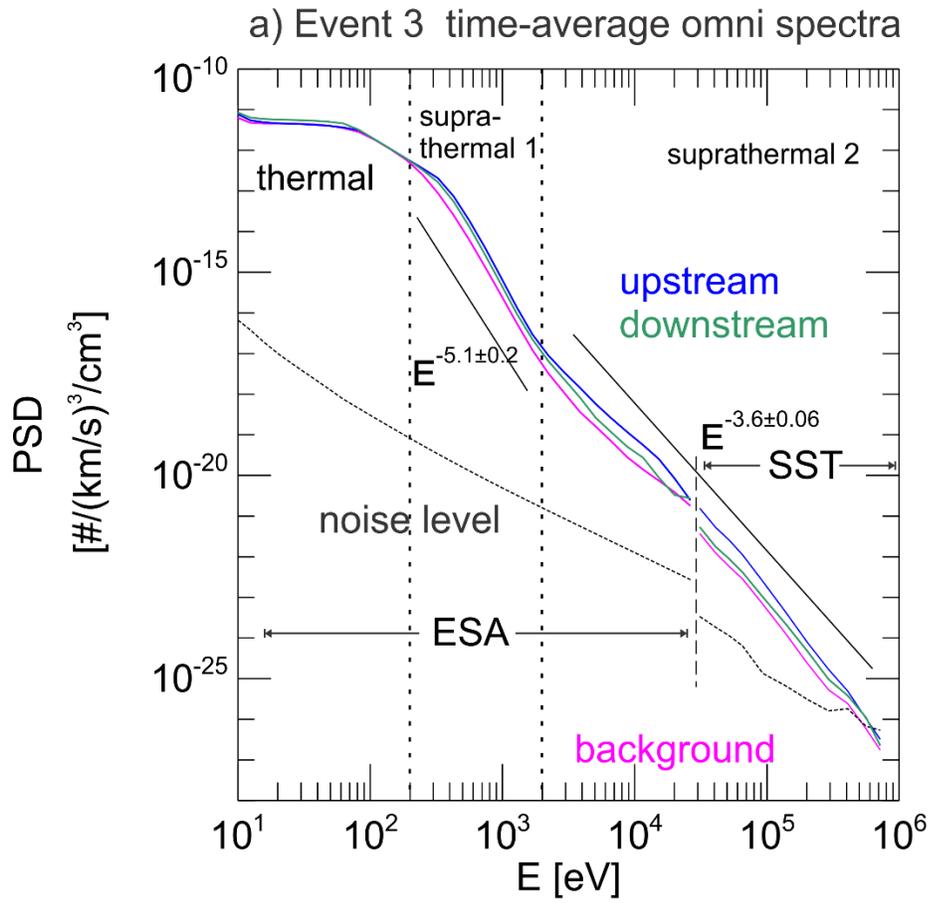

36
37

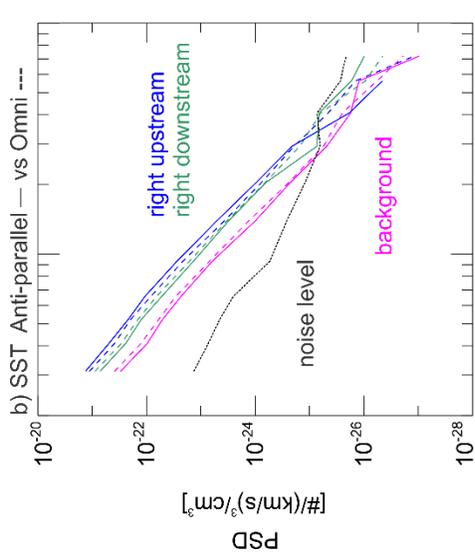
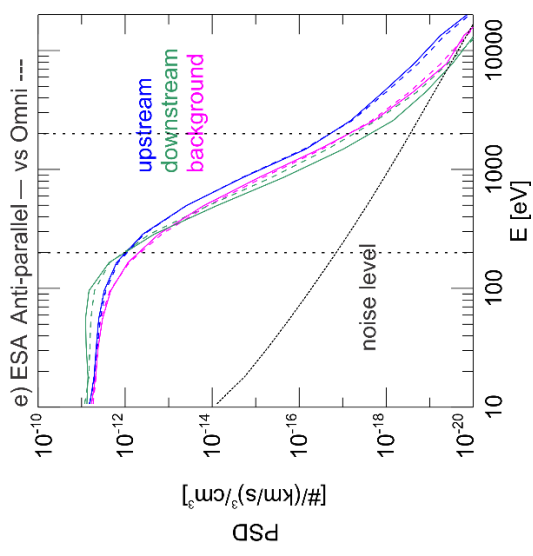
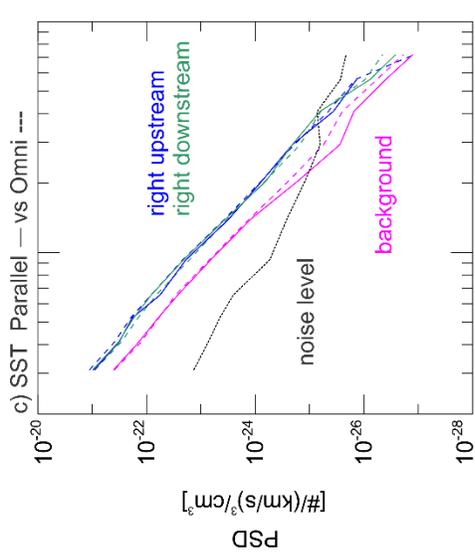
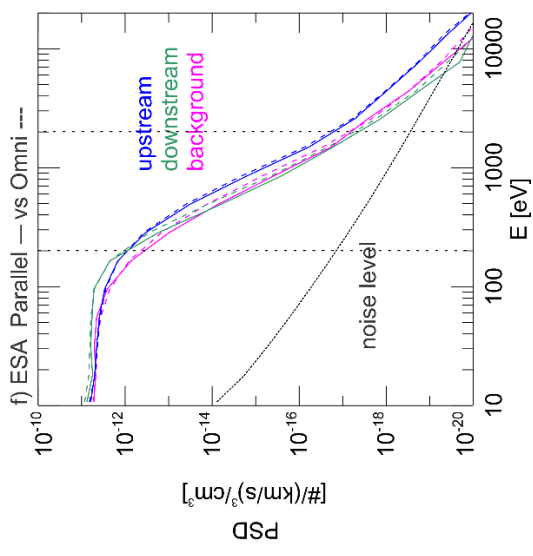
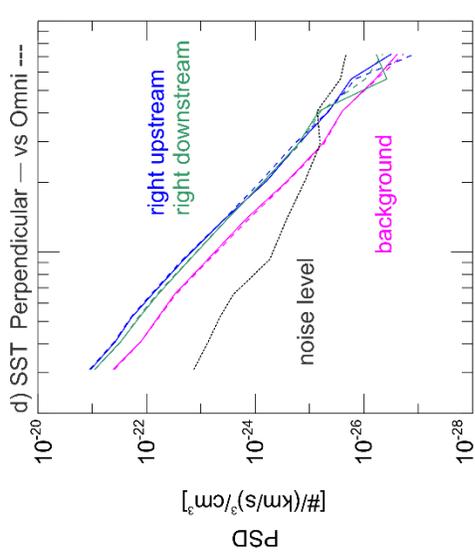
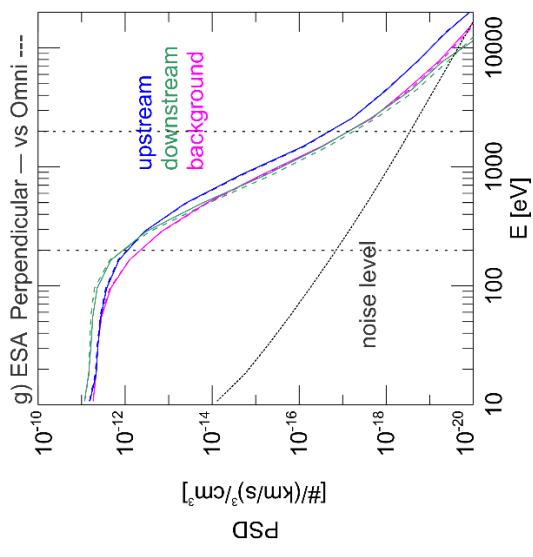

Figure 4. The electron phase space density spectra around the bow wave. (a) the long-time averaged omni-directional electron PSD spectra in the ambient magnetosheath (~17:34-17:36 UT; magenta line) and near the bow wave (~17:36-17:38 UT; blue line). The dotted line is the instrumental noise level. (b)-(d) are the short-time averaged electron PSD measured by SST in the direction anti-parallel, parallel, and perpendicular direction, respectively. Magenta, blue, and green lines are spectra averaged in the ambient magnetosheath (between two magenta lines in Figures 3c-e), right upstream of the bow wave (between vertical blue line and dashed line in Figures 3c-e), and right downstream of the bow wave (between vertical dashed line and green line in Figures 3c-e), respectively. The colored dashed lines are the omni-directional spectra during the same time interval for comparison. (e)-(g) are in the same format as (b)-(d) but measured by ESA. Their time intervals are corresponding to vertical colored lines in Figures 3h-j. The vertical dotted lines indicate 200 eV to 2 keV (suprathermal 1).

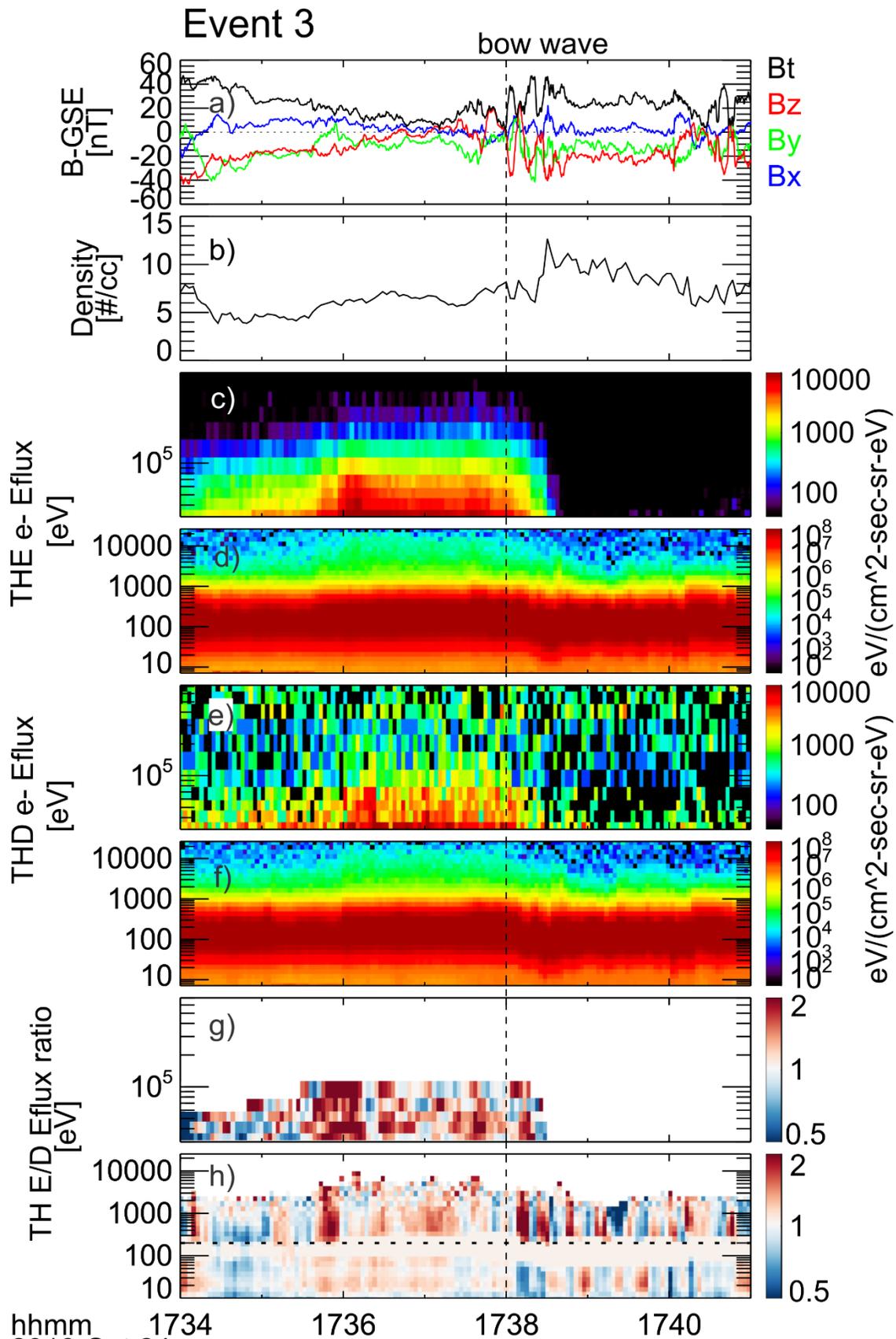
51

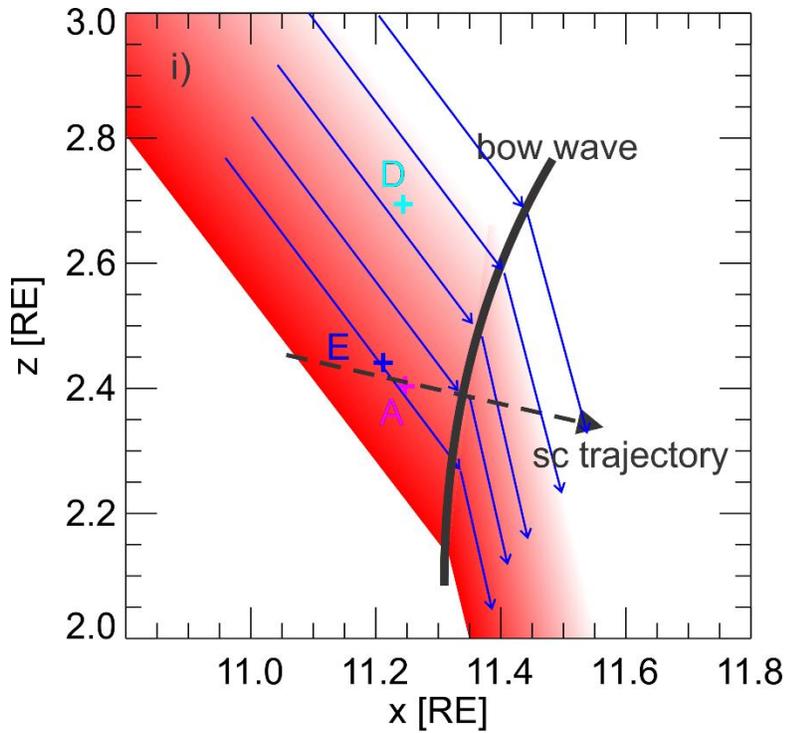

**Figure 5.** The electron energy flux comparison between TH-E and TH-D observations. (a)-(d) are magnetic field, density, electron energy flux spectra observed by TH-E. (e) and (f) are electron energy flux spectra from 7 eV to 25 keV and from 30 keV to 700 keV observed by TH-D. (g) and (h) are the ratio between (c) and (e) and between (d) and (f), respectively. (i) is the zoomed in sketch of Figure 2.

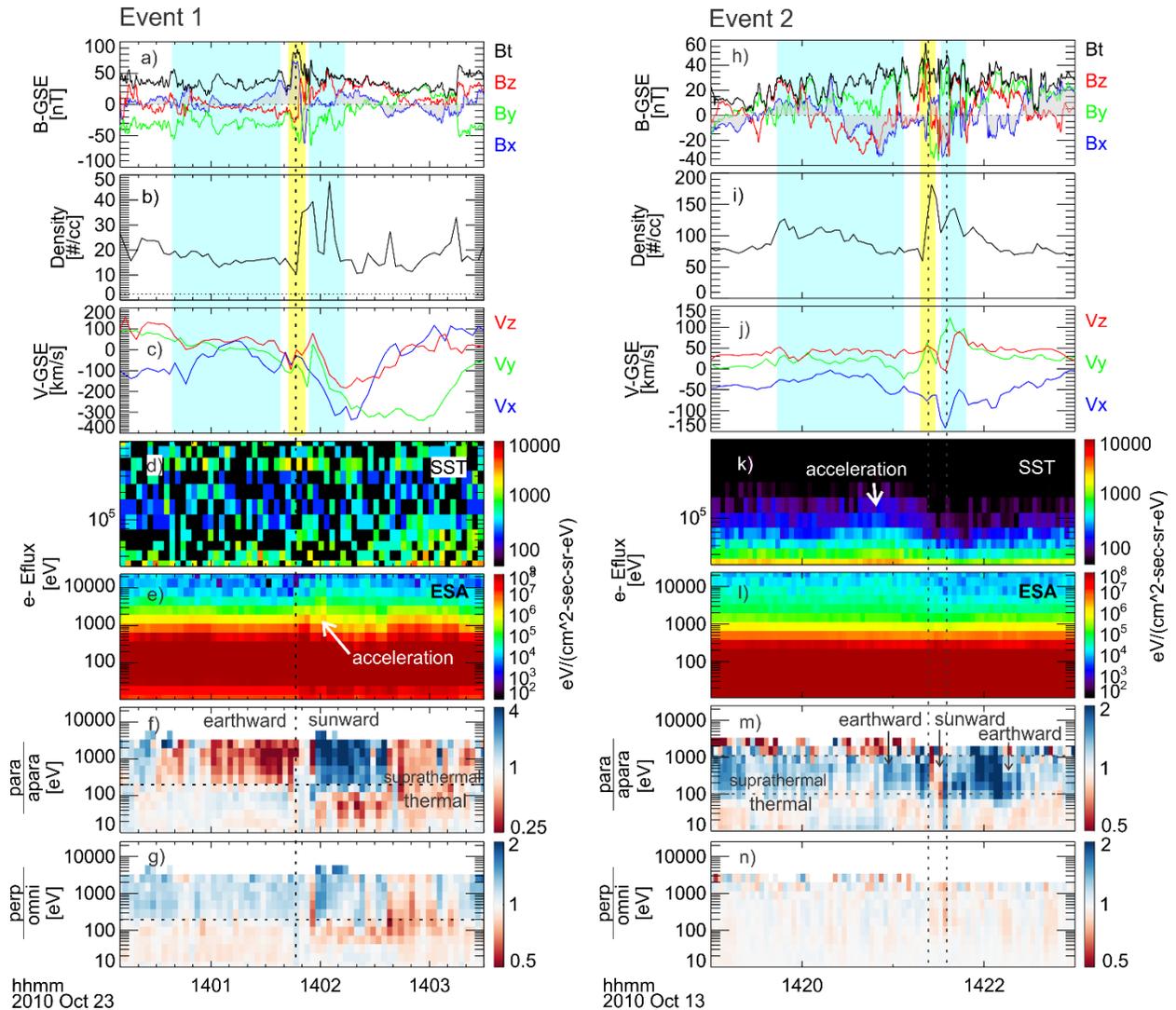

**Figure 6.** The results for event 1 and 2. (a) to (g) are magnetic field in GSE, density, ion bulk velocity in GSE, electron energy flux spectra, and the ratio of parallel flux to anti-parallel flux, the ratio of perpendicular flux to omni-directional flux, respectively. (h) to (n) are the same format as (a) to (g). Blue and yellow regions are the same as in Figures 1.1, 1.2.

**Acknowledgement**

The work at UCLA and SSI was supported by NASA grant NNX17AI45G. TZL is supported by the NASA Living With a Star Jack Eddy Postdoctoral Fellowship Program, administered by the Cooperative Programs for the Advancement of Earth System Science (CPAESS). HH was supported by the Royal Society University Research Fellowship URF\R1\180671 and the Turku Collegium for Science and Medicine. The work in the University of Turku was performed in the framework of the Finnish Centre of Excellence in Research of Sustainable Space. RV acknowledges the financial support of the Academy of Finland (projects 309939 and 312357). We thank the THEMIS software team and NASA's Coordinated Data Analysis Web (CDAWeb, http://cdaweb.gsfc. nasa.gov/) for their analysis tools and data access. The THEMIS data and THEMIS software (TDAS, a SPEDAS v3.1 plugin, see Angelopoulos et al. (2019)) are available at http://themis.ssl.berkeley.edu.